\documentclass[sn-aps,twocolumn]{sn-jnl}

\usepackage{graphicx}
\usepackage{multirow}
\usepackage{amsmath,amssymb,amsfonts}
\usepackage{xcolor}
\usepackage[T1]{fontenc}
\usepackage[utf8]{inputenc}
\usepackage{physics}

\begin{document}

\title[Title]{Hall Coefficient of the Intercalated Graphite CaC$_6$ in the Uniaxial CDW Ground State}

\author[1]{\fnm{Petra} \sur{Đurkas Grozi\'{c}}}

\author[1]{\fnm{Barbara} \sur{Keran}}

\author[1,2]{\fnm{Anatoly M.} \sur{Kadigrobov}}

\author[1]{\fnm{Zoran} \sur{Rukelj}}

\author[1]{\fnm{Ivan} \sur{Kup\v{c}i\'{c}}}

\author*[1]{\fnm{Danko} \sur{Radi\'{c}}}\email{dradic@phy.hr}

\affil[1]{\orgdiv{Department of Physics}, \orgname{University of Zagreb Faculty of Science}, \orgaddress{\street{Bijeni\v{c}ka 32}, \city{Zagreb}, \postcode{10000}, \country{Croatia}}}

\affil[2]{\orgdiv{Theoretische Physik III}, \orgname{Ruhr-Universit\"{a}t Bochum}, \orgaddress{\street{Universit\"{a}tsstrasse 150}, \city{Bochum}, \postcode{D-44801}, \country{Germany}}}

\abstract{We evaluate the Hall coefficient characterising magnetotransport in an intercalated graphite CaC$_6$ with the Fermi surface reconstructed by an uniaxial charge density wave from closed pockets to open sheets. As the typical order parameter, corresponding to the pseudo-gap in electronic spectrum and consequently to spacing between electron trajectories in reciprocal space, is of the order of $10^2$K, magnetic breakdown in strong experimentally achievable fields of the order of 10T is inevitable. The classical expressions for the components of the magnetoconductivity tensor are strongly modified by magnetic field-assisted over-gap tunneling causing quantum interference. Due to magnetic breakdown, all magnetoconductivity components undergo strong quantum oscillations reflected in the Hall coefficient. In their nature, these are different than standard Shubnikov de Haas oscillations which would not appear in a system with an open Fermi surface.}


\maketitle

\section{Introduction}

The Hall coefficient is a parameter that facilitates the understanding of the electrical properties of materials in the presence of magnetic fields. It provides insight into the behaviour of charge carriers in a material and has significant implications for various physical phenomena, including magnetotransport properties and the topology of the Fermi surface (FS).  In that sense, our study is focused on the intercalated graphite compounds (GIC), particularly on CaC$_6$. Even though CaC$_6$ is mostly studied for its superconducting properties \cite{Dresselhaus}, our main focus in this work is its behaviour in the uniaxial charge density wave (CDW) ground state \cite{Rahnejat}.
In materials where the CDW ground state consequently imposes the FS reconstruction \cite{Petra}, a strong magnetic field may induce the over-gap tunnelling of electrons, so-called magnetic breakdown (MB), whose signature we want to detect in the Hall coefficient.  

In this paper,  we present a theoretical derivation of the Hall coefficient under MB conditions. It is based on the calculation of the magnetoconductivity tensor components using the quantum density matrix formalism within the semiclassical approximation.
The striking feature of magnetic breakdown is relating the huge semiclassical phases, i.e. action integrals, to interference, thus transforming the classical result to non-classical. It results in an onset of quantum oscillations in the system with open FS where one would not expect them in terms of standard Shubnikov - de Haas effect.

The paper is organised as follows: In Section 2, we present the model of the CaC$_6$ system and derive the dispersion law under the conditions of magnetic breakdown; in Section 3,  we present the derivation of the magnetoconductivity tensor and Hall coefficient,  followed by results; in Section 4, we present the effective carrier concentrations participating in DC conductivity and magnetoconductivity along different directions; in the Conclusions, we present discussion and concluding remarks.\\

\section{The model and theoretical framework}

\begin{figure}
\centering
\includegraphics[width=0.9\columnwidth]{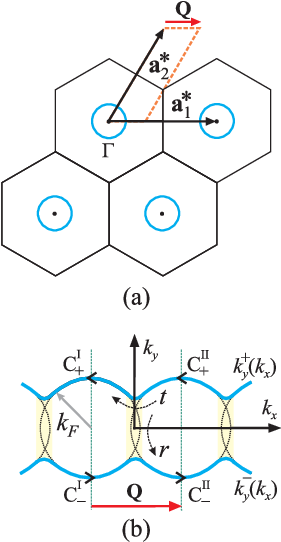}
\caption{A schematic representation of a 2D layer of CaC$_6$ in reciprocal space. 
(a) The CaC$_6$ Brillouin zone (black hexagon), with reciprocal unit vectors $\mathbf{a}_{1,2}^*$ ($\vert \mathbf{a}^*_{1,2}\vert\approx 1.68$\AA$^{-1}$). The Fermi surface is approximated by a circle at the $\Gamma$ point. Chemical doping of 0.2 electrons per carbon atom \cite{Rahnejat} results in the Fermi pocket with an area of $S_0 \approx 0.244$\AA$^{-2}$ and average Fermi wave number $k_F \approx 0.28$\AA$^{-1}$. The CDW potential, with the wave vector $\mathbf{Q}\| \mathbf{a}_1^*$ of periodicity $Q=|\mathbf{a}^*_1|/3$ reduces the CaC$_6$ BZ three times along that direction (dashed orange rhombus), bringing the Fermi pockets into contact where the degeneracy is lifted by gap opening.
(b) The CDW potential reconstructs the FS into open sheets along the $k_x$ direction. Arrows indicate the semiclassical motion of electrons in an external magnetic field $\mathbf{B}$ perpendicular to the sample. Electrons pass through the MB-junction (shaded) with probability amplitude $t$, or are reflected from it with probability amplitude $r$ due to magnetic breakdown. The coefficients $C_\pm^{\mathrm{I,II}}$ denote the branches of semiclassical wave functions that correspond to trajectories $k_{y}^{\pm}(k_x)$ (arrows) on the left (I) / right (II) side of the MB node in the origin.}
\label{figMB}
\end{figure}

The CaC$_6$ material is modelled as a 2D graphene sheet doped with electrons from intercalated Ca atoms, creating an electron pocket at the Fermi energy. The hexagonal Ca superlattice triples the size of the carbon primitive cell, folding the original carbon Brillouin zone (BZ) to a new one three times smaller.
The resulting FS can be approximated as a 6-fold degenerate circle in the $\Gamma$-point of the CaC$_6$ BZ (degeneracy factor $\zeta=6$ comes form 2-fold graphene valley degeneracy and trippling of the unit cell), with details of the Fermi pocket shape incorporated as parameters in the effective carrier concentrations participating in conductivity calculations. A uniaxial CDW forms with peaks along the graphene armchair direction, tripling the CaC$_6$ cell in one direction. This further BZ reduction causes Fermi pockets to touch (or slightly overlap), leading to FS reconstruction due to the CDW order parameter acting as a gap parameter $\Delta$. This transforms the FS from the form of closed Fermi pockets into open contours. To study magnetoconductivity, a homogeneous magnetic field $\mathbf{B}$ is applied perpendicular to the sample plane. Schematic presentation of reciprocal space configuration is shown in Fig. \ref{figMB}(a).\\
%
%

To determine the electron spectrum under MB conditions, a semiclassical approach based on the Lifshitz-Onsager Hamiltonian \cite{Onsager,Lifshitz} is employed. This method describes the semiclassical electron motion between MB regions shown schematically in Fig. \ref{figMB}(b). This approach assumes a coherent magnetic breakdown which appears at sufficiently strong magnetic field, resulting in a Larmor radius much smaller than the mean free path for impurity scattering, and the absence of dislocation fields.\\

With the Landau gauge choice for the vector potential $\mathbf{A} = (0, Bx, 0)$, the Lifshitz-Onsager Hamiltonian yields a Schrödinger equation in reciprocal space
%
\begin{align}
\varepsilon_{\nu}\left(k_x, K_y - i\frac{b_B^2}{\hbar^2}\frac{\mathrm{d}}{\mathrm{d}k_x} \right) G_{\nu}(k_x, K_y) \nonumber \\
=\varepsilon \, G_{\nu}(k_x,K_y).
\label{Schrodinger}
\end{align}
%
There, $\varepsilon_\nu (\mathbf{k})=\hbar v_F \vert\mathbf{k} \vert$ is the initial electron spectrum of the unreconstructed band with the Fermi velocity $v_F$ and 2D electron wave vector $\mathbf{k}=(k_x,k_y)$, index $\nu$ accounting for the $\pm$ branch in the I, II region (see Fig. \ref{figMB}(b)). Parameter $b_B=\sqrt{e\hbar B}$ is the "magnetic length" in $k$-space for an electron with charge $-e$, and $\hbar K_y$ is the conserved generalised momentum of the semiclassical motion of the electron in the used gauge. 
We solve Eq. (\ref{Schrodinger}) using the semiclassical wave functions
%
\begin{align}
& G_{\pm}(k_x, K_y)=  \frac{C_{\pm}}{\sqrt{\vert v_y^{\pm}\vert}} \nonumber\\
& \times\exp\left[i\frac{\hbar^2}{b_B^2}\int^{k_x}\left(k_y^{\pm}(k_x';\varepsilon_F)-K_y\right)\mathrm{d}k_x'\right]
\label{WF}
\end{align}
%
that are analogous for both regions I and II, with the corresponding coefficients $C_{\pm}$, group velocities $v^{\pm}_y$ (i.e. $\mathbf{v}=\tfrac{1}{\hbar}\nabla_{\mathbf{k}} \varepsilon(\mathbf{k})$) and $k_y^{\pm}( k_x;\varepsilon)=\pm \sqrt{(\varepsilon/\hbar v_F)^2 - k_x^2}$.
The semiclassical solutions $G_\pm^{\mathrm{I,II}}$ are connected by the MB junction, which is described by so-called MB matrix
%
\begin{eqnarray}
\begin{pmatrix}
C_-^\mathrm{I}\\
C_+^\mathrm{II}
\end{pmatrix}=
\mathrm{e}^{i\theta}
\begin{pmatrix}
t & r\\
-r^{\ast} & t^{\ast}
\end{pmatrix}
\begin{pmatrix}
C_-^\mathrm{II}\\
C_+^\mathrm{I}
\end{pmatrix},
\label{MB_matrix}
\end{eqnarray}
%
where $t(B)$ and $r(B)$ are the complex probability amplitudes for an electron to pass through the MB junction or to be reflected on it. The corresponding probabilities satisfy the unitarity condition $\vert t\vert^2 + \vert r\vert^2=1$.
It was shown \cite{EPJB,PhysicaB,Voroncov,Fortin} that in the situation where the FS reconstruction takes place in the way that the initial Fermi pockets nearly touch or very slightly overlap, the transmission probability is of the form 
%
\begin{eqnarray}
|t(B)|^2\approx 1- \exp \left[-\frac{\Delta^2}{\hbar \omega_c\varepsilon_F} \sqrt[3]{\frac{\varepsilon_F}{\hbar \omega_c}} \, \right],
\label{MBprobab-t}
\end{eqnarray}
%
where $\hbar \omega_c$ is magnetic energy, $\omega_c=\tfrac{eB}{m^*}$ is cyclotron frequency, $m^*$ is electron effective cyclotron mass.
Imposing periodic boundary conditions upon the semiclassical solutions, i.e. $G_{\pm}(k_x, K_y)=G_{\pm}(k_x+Q, K_y)$, gives us two additional relations between four coefficients $C_\pm^{\mathrm{I,II}}$. All together we obtain a system of two algebraic equations for two unknowns with determinant at energy $\varepsilon$,
%
\begin{eqnarray}
D(\varepsilon, K_y) &=& \cos \left( \frac{ \hbar^2 S_0(\varepsilon)}{2 b_B^2} + \theta \right) \nonumber\\
&-& |t|\cos\left( \frac{ \hbar^2 Q K_y}{b_B^2} +\mu \right),
\label{D}
\end{eqnarray}
%
where $S_0(\varepsilon) = \int_0^Q k_y^+(k_x;\varepsilon)dk_x = \int_Q^0 k_y^-(k_x;\varepsilon)dk_x$ is area of reciprocal space enclosed by electron trajectory (dotted circle in Fig. \ref{figMB}(b)) and $\theta$, $\mu$ are phases acquired from the boundary condition in particular MB configuration \cite{Petra2}. Spectrum of the problem is determined by the dispersion law $D(\varepsilon, K_y)=0$.

\section{The Hall coefficient}

The Hall coefficient, by the definition and expressed within the linear response theory, reads  
\begin{align}
R_H \equiv \frac{E_y}{j_x B}=\frac{\sigma_{xy}}{(\sigma_{xx}\sigma_{yy}+\sigma_{xy}^2)B}
\label{HallCoefDef}
\end{align}
where $\sigma_{ij}$ are components of the magnetoconductivity tensor {\boldmath $\sigma$}, relating electric current $\mathbf{j}$ and electric field $\mathbf{E}$ in the presence of magnetic field $\mathbf{B}$ perpendicular to the $(x,y)$-plane, i.e. $\mathbf{j}=${\boldmath $\sigma$}$(\mathbf{B})\mathbf{E}$ \cite{Abrikosov}. 
To derive the magnetoconductivity tensor in the magnetic breakdown regime, we employ the quantum density matrix formalism. The equation of motion for the total density matrix of the system $\hat{\rho}_{tot}$ in the relaxation time approximation is given by expression
\begin{align}
\dot{\hat{\rho}}_{tot} +\frac{\hat{\rho}_{tot}-f_0(H_0)}{\tau}=0,
\label{EqOfMotion_0}
\end{align}
where $f_0(H_0)$ is the equilibrium Fermi-Dirac distribution, $H_0$ is the unperturbed Hamiltonian satisfying $[f_0, H_0]=0$, and $\tau$ is the relaxation time of a system.
The first  term in Eq. (\ref{EqOfMotion_1}) can be substituted using the Heisenberg equation of motion for the total Hamiltonian $H$, i.e. $i\hbar \dot{\hat{\rho}}_{tot}= \left[\hat{\rho}_{tot}, H \right]$, while the total density matrix can be presented as $\hat{\rho}_{tot}=f_0 + \hat{\rho}$, $ \hat{\rho}$ being a nonequilibrium correction within the linear response theory. Plugging in the total Hamiltonian $H=H_0 - eV(\mathbf{r})$, where $V(\mathbf{r})$ is the perturbation potential imposed by an external electric field acting on electron with charge $-e$, and keeping only the linear terms in the expansion, we reduce Eq. (\ref{EqOfMotion_0}) to
\begin{align}
-\frac{i}{\hbar}\left[\hat{\rho}, H_0 \right] + \frac{\hat{\rho}}{\tau}=-i\frac{e}{\hbar}\left[f_0, V(\mathbf{r}) \right].
\label{EqOfMotion_1}
\end{align}
%
Taking the $\bra{\eta}\cdots\ket{\eta'}$ matrix elements of the commutators in Eq. (\ref{EqOfMotion_1}), where $\ket{\eta}=\Psi_{n,\mathbf{K}}(\mathbf{r})$ is electron wave function in $n$-th Landau band with the wave vector $\mathbf{K}$, we obtain
\begin{align}
&\left[ \frac{i}{\hbar} \left( \varepsilon_{\eta}-\varepsilon_{\eta'} \right) +\frac{1}{\tau}\right] \rho_{\eta\eta'}= \nonumber\\
&-i\frac{e}{\hbar}\left(f_0(\varepsilon_{\eta})-f_0(\varepsilon_{\eta'})\right)\bra{\eta}V(\mathbf{r})\ket{\eta'}.
\label{eq.ro1}
\end{align}
%
To evaluate the matrix element $\bra{\eta}V(\mathbf{r})\ket{\eta'}$, we shift its coordinate variable $\mathbf{r}\rightarrow \mathbf{r} + \mathbf{a}$, where $\mathbf{a}=a_j\hat{\mathbf{e}}_j$ is the periodicity vector in the 2D crystal lattice defined by the $\hat{\mathbf{e}}_j$, $j\in\{x,y\}$ unit vectors, by which $\Psi_{\eta}(\mathbf{r+a})=\exp(i\mathbf{K\cdot a})\Psi_{\eta}(\mathbf{r})$. Then we expand $V(\mathbf{r+a})$ into the Taylor series in $\mathbf{a}$, i.e.
\begin{small}
\begin{align}
\bra{\eta}V(\mathbf{r})\ket{\eta'} = \int  \Psi_{n,\mathbf{K}}^{*}(\mathbf{r}) V(\mathbf{r})\Psi_{n',\mathbf{K}'}(\mathbf{r})\mathrm{d}^{3}r \nonumber\\
 \approx \mathrm{e}^{-i(\mathbf{K}-\mathbf{K}')\mathbf{a}}\bra{\eta}V(\mathbf{r})\ket{\eta'} \nonumber\\
 + \int  \Psi_{n,\mathbf{K}}^{*}(\mathbf{r})\mathbf{\nabla}V(\mathbf{r}) \cdot \mathbf{a} \, \Psi_{n',\mathbf{K}'}(\mathbf{r})\mathrm{d}^{3}r. \nonumber\\
\end{align}
\end{small}
Inserting the electric field $\mathbf{E}=-\nabla V(\mathbf{r})$ and expanding the exponential function $\exp((\mathbf{K-K}')\cdot \mathbf{a})$ up to the linear term, we obtain the matrix element 
\begin{align}
\bra{\eta}V(\mathbf{r})\ket{\eta'}=\mathbf{E}\frac{\mathbf{a}}{i(\mathbf{K}-\mathbf{K'})\cdot \mathbf{a}}\bra{\eta}\ket{\eta'}.
\label{matricnielement_V}
\end{align}
%
On the other hand, starting from the Heisenberg equation for the velocity operator, i.e. $i\hbar \hat{\mathbf{v}}= \left[\hat{\mathbf{r}}, H_0 \right]$, using the same procedure of shifting the coordinate and Taylor expansion to evaluate matrix elements of commutator on the right-hand side, we obtain 
\begin{align}
\bra{\eta}\hat{\mathbf{v}}\ket{\eta'}=-\frac{(\varepsilon_{\eta}-\varepsilon_{\eta'})\mathbf{a}}{(\mathbf{K}-\mathbf{K'})\cdot \mathbf{a}}\bra{\eta}\ket{\eta'}.
\label{matricnielement_v}
\end{align}
Plugging expression (\ref{matricnielement_v}) into (\ref{matricnielement_V}) and then into Eq. (\ref{eq.ro1}) we obtain the density matrix element
\begin{align}
\rho_{\eta\eta'}=e\mathbf{E} \frac{\bra{\eta}\hat{\mathbf{v}}\ket{\eta'}}{\tfrac{i}{\hbar}(\varepsilon_\eta-\varepsilon_\eta')+\tfrac{1}{\tau}}\frac{f_0(\varepsilon_{\eta})-f_0(\varepsilon_{\eta'})}{\varepsilon_{\eta}- \varepsilon_{\eta'}}.
\label{eq.ro2}
\end{align}

Knowing the density matrix we can calcualte the 2D current density, $\mathbf{j} = -2\zeta e \tfrac{1}{A} \mathrm{Tr} \left[ \hat{\rho} \hat{\mathbf{v}} \right] = -2\zeta e \tfrac{1}{A} \sum_{\eta\eta'} \bra{\eta} \hat{\rho} \ket{\eta'}  \bra{\eta'} \hat{\mathbf{v}} \ket{\eta}$ where factor 2 appears due to spin degeneracy, $\zeta=6$ appears due to graphene double valley degeneracy and tripling the graphene unit cell by Ca superlattice, $A=L_x \times L_y$ is area of the sample. Expressing the inner product $\mathbf{E}\cdot \bra{\eta}\hat{\mathbf{v}}\ket{\eta'}=\sum_\beta E_\beta \bra{\eta}\hat{v}_\beta\ket{\eta'}$ and $\bra{\eta'}\hat{\mathbf{v}}\ket{\eta}=\sum_\alpha \bra{\eta'}\hat{v}_\alpha\ket{\eta} \hat{\mathbf{e}}_\alpha$, using Eq. (\ref{eq.ro2}), we can express the current component
\begin{align}
j_\alpha=-\frac{2\zeta e^2}{L_xL_y}\sum_\beta \sum_{\eta\eta'}\frac{\bra{\eta}v_\beta\ket{\eta'}\bra{\eta'}v_\alpha\ket{\eta}}{\tfrac{i}{\hbar}(\varepsilon_{\eta}-\varepsilon_{\eta'})+\tfrac{1}{\tau}} \nonumber\\
\times \frac{f_0(\varepsilon_{\eta})-f_0(\varepsilon_{\eta'})}{\varepsilon_{\eta}- \varepsilon_{\eta'}} E_\beta.
\label{struja_1}
\end{align}
Utilising the afore-mentioned definition of conductivity tensor, i.e. $j_\alpha = \sum_\beta \sigma_{\alpha \beta} E_\beta$, we can extract the conductivity tensor from Eq. (\ref{struja_1})
\begin{align}
\sigma_{\alpha \beta}=-\frac{2\zeta e^2}{L_xL_y} \sum_{\eta\eta'}\frac{\bra{\eta}v_\beta\ket{\eta'}\bra{\eta'}v_\alpha\ket{\eta}}{\tfrac{i}{\hbar}(\varepsilon_{\eta}-\varepsilon_{\eta'})+\tfrac{1}{\tau}} \nonumber\\
\times \frac{f_0(\varepsilon_{\eta})-f_0(\varepsilon_{\eta'})}{\varepsilon_{\eta}- \varepsilon_{\eta'}}.
\label{vodljivost_general}
\end{align}
The difference of Fermi functions divided by difference of energies can be approximated by derivative if the Fermi function is smooth enough along that energy scale, i.e. temperature is high enough, yielding the expression
\begin{align}
\sigma_{\alpha \beta}=-\frac{2\zeta e^2}{L_xL_y} \sum_{\eta\eta'}\frac{\bra{\eta}v_\beta\ket{\eta'}\bra{\eta'}v_\alpha\ket{\eta}}{\tfrac{i}{\hbar}(\varepsilon_{\eta}-\varepsilon_{\eta'})+\tfrac{1}{\tau}} \nonumber\\
\times \frac{\mathrm{d}f_0}{\mathrm{d}\varepsilon}\Big|_{\varepsilon=\varepsilon_{\eta}}
\label{vodljivost}
\end{align}
which we shall use in this paper.\\

The details of calculation of magnetoconductivity components $\sigma_{xx}$, $\sigma_{yy}$ and $\sigma_{xy}$ using the semiclassical wave functions (\ref{WF}) under the MB conditions are presented in Ref. \cite{Petra2}. There, the influence of the MB on the main, classical contribution to magnetoconductivity was studied, neglecting the additive corrections of higher order. The components of magnetoconductivity tensor at temperature much smaller comparing to $\hbar \omega_c$ scale (technically the $T=0$ limit), $xx$ being along direction perpendicular to open electron trajectories in reciprocal space, $yy$ along direction of the open trajectories, and $xy$ being the Hall conductivity, are
\begin{align}
\sigma_{xx}=&\frac{m^* n_0}{\tau B^2} \mathcal{O}(B) \nonumber\\
\sigma_{yy}=& \frac{e^2 \tau \zeta Q^2}{\pi^3 m^*}  \mathcal{O}^{-1}(B) \nonumber\\
\sigma_{xy}=& -\frac{en_0}{B} \mathcal{O}(B)
\label{sigma_components}
\end{align}
in which the oscillating function is
\begin{small}
\begin{align}
&\mathcal{O}(B) = \nonumber\\
&\frac{\Big| \sin \left( \frac{\pi}{2}\frac{\varepsilon_F}{\hbar\omega_c}\right)\Big|}{\sqrt{\vert t\vert^2- \cos^2\left( \frac{\pi}{2}\frac{\varepsilon_F}{\hbar\omega_c}\right) }} \Theta\left[\vert t \vert^2 - \cos^2\left( \frac{\pi}{2}\frac{\varepsilon_F}{\hbar\omega_c}\right)  \right] \nonumber\\
\label{oscillatory_functions}
\end{align}
\end{small}
where $n_0=2\zeta S_0/(2\pi)^2$ is the surface carrier concentration related to area of the Fermi pocket $S_0$.\\

\begin{figure}
\centering
\includegraphics[width=\columnwidth]{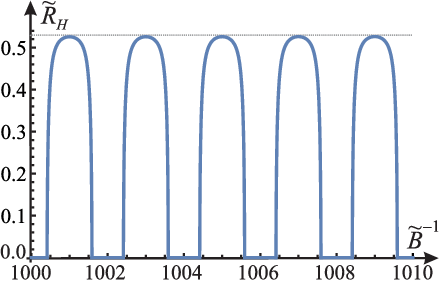}
\caption{Oscillations of the zero-temperature Hall coefficient $R_H$ vs. inverse magnetic field $B^{-1}$. Values are scaled as $\widetilde{R}_H \equiv R_H / (-1/en_0)$ and $\widetilde{B} \equiv \hbar \omega_c / \epsilon_F$. Period of oscillations is proportional to area of the FS $S_0$. Presented region corresponds to the field value around 10T with $|t(B)|^2 \approx 0.632$. Dotted line is an envelope of oscillations which in the low field with the MB absent (i.e. $|t(B)|^2 \rightarrow 1$)  saturates to value $(1+8/\pi^2)^{-1}\approx 0.552$.}
\label{fig1_RH0}
\end{figure}

Finally, we obtain the zero temperature Hall coefficient
\begin{align}
R_H=-\frac{1}{en_0} \, \frac{\mathcal{O}(B)}{\frac{\zeta Q^2}{\pi^3n_0}+\mathcal{O}^2(B)}.
\label{Hallcoef}
\end{align}
The numerical term in the numerator of expression (\ref{Hallcoef}) is determined by the properties of the system which, taking $S_0 \approx \pi k_F^2$, $Q \approx 2k_F$ and $\zeta=6$, evaluates to $\zeta Q^2/(\pi^3 n_0) = 8/\pi^2$.
In the absence of MB in the low field limit when $|t(B)^2|\rightarrow 1$, we find $\mathcal{O}(B) \approx 1$ yielding the Hall coefficient
\begin{align}
R_H=-\frac{1}{en_0}\frac{1}{1+\frac{8}{\pi^2}}
\label{Hallcoef_t=1}
\end{align}
being a (negative) constant modified from the standard value by the factor $(1+8/\pi^2)^{-1}$.
In the presence of finite MB, i.e. when $|t(B)^2| < 1$ (see Eq. (\ref{MBprobab-t}), the Hall coefficient exhibits oscillations periodic with $B^{-1}$ as shown in Fig. \ref{fig1_RH0}.
%
%
Analysis of Eq. (\ref{vodljivost}) together with numerical evaluation at finite temperatures \cite{Petra2} show that oscillations are more pronounced the lower is the temperature and they persist up to the temperature scale of the order of $\hbar \omega_c$ above which they get "smoothed out" exponentially in the similar manner as the standard Shubnikov - de Haas oscillations appearing in systems with closed FS.\\

\section{Effective charge carrier densities}


In expressions (\ref{sigma_components}) for components of magnetoconductivity tensor there appears the electron surface concentration $n_0$. It is determined by the area of the FS $S_0=\pi k_F^2$, assuming its isotropic circular shape with band dispersion $\varepsilon_\mathbf{k}=\hbar v_F |\mathbf{k}|$, and evaluating to $n_0=\zeta \varepsilon_F^2/(2\pi\hbar^2 v_F^2)$. This assumption is approximative, neglecting the effects of reconstruction of the FS turned from closed to open contours, i.e. its anisotropy. That taken into account, one can define the effective concentrations of carriers taking part in DC electric transport and magnetotransport along a particular direction: $n_{xx}$ - along the reconstruction direction $\mathbf{Q}$, $n_{yy}$ - perpendicular to the reconstruction direction, $n_H$ - Hall concentration. Here we present the standard zero-field values in the absence of magnetic breakdown. Calculation of concentrations in the regime of finite breakdown is beyond the scope of this paper.\\

\begin{figure}
\centering
\includegraphics[width=\columnwidth]{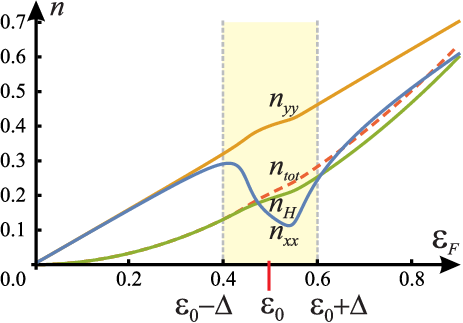}
\caption{Effective concentrations of carriers vs. Fermi energy: $n_{xx}$ - along the reconstruction direction $\mathbf{Q}$, $n_{yy}$ - perpendicular to the reconstruction direction, $n_H$ - Hall concentration, $n_{tot}$ - total concentration (closest to $n_0 \sim \varepsilon_F^2$), all calculated at $T=70$K. $\varepsilon_0$ is the middle of the pseudo-gap (shaded interval) appearing due to the CDW with order parameter $\Delta$. Energy is scaled to $\hbar v_F Q$, and concentration to $3m_e v_F Q / \pi^2 \hbar$.}
\label{fig_koncentracije}
\end{figure}

To obtain the effective concentrations, first we define so-called dimensionless reciprocal effective mass tensor 
%
\begin{eqnarray}
 M_{\alpha \beta} (\mathbf{k}) = \frac{m}{\hbar^2} \frac{\partial^2 \epsilon_{\mathbf{k}}}{\partial k_{\alpha} \partial k_{\beta}}
\label{M_ij}
\end{eqnarray}
where $m$ is the bare electron mass and $\epsilon_{\mathbf{k}}$ is the reconstructed electron band dispersion of the form
%
\begin{small}
\begin{align}
&\epsilon_{\pm}(\mathbf{k})= \frac{1}{2} \left[ \varepsilon (\mathbf{k}-\tfrac{\mathbf{Q}}{2})+\varepsilon ( \mathbf{k}+\tfrac{\mathbf{Q}}{2} ) \right. \nonumber\\
&\left. \pm
 \sqrt{\left( \varepsilon ( \mathbf{k}-\tfrac{\mathbf{Q}}{2} )-\varepsilon ( \mathbf{k}+\tfrac{\mathbf{Q}}{2} ) \right)^2+4\Delta^2} \right]. \nonumber\\
\label{Bands}
\end{align}
\end{small}
%
The total concentration of electrons is defined as 
%
\begin{eqnarray}
n_{tot} = \frac{2\zeta}{A} \sum_{\mathbf{k}} f_0(\epsilon_\mathbf{k}).
\label{n_tot}
\end{eqnarray}
%
Effective concentration of electrons $n_{\alpha \alpha} \equiv -\tfrac{2\zeta}{A} \sum_{\mathbf{k}} m v_{\alpha \mathbf{k}}v_{\alpha \mathbf{k}} \tfrac{\mathrm{d} f_0}{\mathrm{d} \epsilon_{\mathbf{k}}}$, which depends on the direction $\alpha$ with respect to the reconstruction wave vector $\mathbf{Q}$, is defined in terms of group velocities $v_{\alpha \mathbf{k}} = (1/\hbar)\partial \epsilon_\mathbf{k} / \partial k_\alpha$. Using expression (\ref{M_ij}) we can express it as
%
\begin{eqnarray}
n_{\alpha \alpha} = \frac{2\zeta}{A} \sum_{\mathbf{k}}  M_{\alpha \alpha} (\mathbf{k}) f_0(\epsilon_{\mathbf{k}}).
\label{n_ii}
\end{eqnarray}
%
The Hall concentration $n_H$ which defines the Hall coefficient in (vanishing) magnetic field perpendicular to the sample plane is defined as  
%
\begin{eqnarray}
n_H = \frac{ n_{xx}n_{yy}}{n_{xy}},
\label{n_H} 
\end{eqnarray}
%
where  
%
\begin{small}
\begin{align}
n_{xy}= -\frac{2\zeta}{A} \sum_{\mathbf{k}}  m \left(v_{x \mathbf{k}} v_{y \mathbf{k}} M_{xy} (\mathbf{k}) \right. \nonumber\\
\left. - v_{x \mathbf{k}} v_{x \mathbf{k}} M_{yy} (\mathbf{k}) \right) \frac{\mathrm{d} f_0}{\mathrm{d} \epsilon_{\mathbf{k}}}.
\label{d5} 
\end{align}
\end{small}
%
%
Numerically evaluated effective concentrations are shown in Fig. \ref{fig_koncentracije}. In evaluation, up to the bottom of the upper band, only the lower band $\epsilon_-(\mathbf{k})$ is included in calculation, while above that energy, also $\epsilon_+(\mathbf{k})$ should be takin into account.
%
%

\section{Conclusions}

In summary, we study the magnetotransport in CaC$_6$ system in the uniaxial charge density wave ground state, in which closed pockets comprising the Fermi surface are transformed into open contours, with magnetic field perpendicular to the 2D sample. The order parameter of the pseudo-gap appearing in the spectrum, of the order of $10^2$ K, makes magnetic breakdown effect inevitable in strong magnetic field of the order of 10 T achievable in experiment. We focused our study on the Hall coefficient, being the main signature that characterises magnetotransport, using the semiclassical approximation and density matrix formalism. We find that the expected low temperature (comparing to magnetic energy $\hbar \omega_c$) classical result of $R_H = -1/en_H
$ in the absence of magnetic breakdown at low field, develops strong oscillations from zero to its maximal value in the regime when magnetic breakdown is present. Oscillations are periodic in inverse magnetic field with period determined by the size of the Fermi surface inside the reconstructed Brillouin zone by the CDW, in the similar manner as Shubnikov - de Haas oscillations in systems with closed Fermi surface. SdH oscillations would appear in magnetoconductivity of the CaC$_6$ not undergoing the CDW instability, due to closed Fermi pockets, however, they would appear as a small additive correction comparing to huge oscillations appearing inherently in the classical part due to magnetic breakdown. We have to emphasise that these oscillations are not related with mechanism leading to Shubnikov - de Haas effect, which is not present in the system with open Fermi surface, but are result of  quantum interference in electronic wave function imposed by magnetic breakdown.


\section*{Declarations}

This work was supported by the QuantiXLie Centre of Excellence, a project co-financed by the Croatian Government and European Union through the European Regional Development Fund - the Competitiveness and Cohesion Operational Programme (Grant PK.1.1.02).

\end{document}